\documentclass[fleqn,10pt]{wlscirep}
\usepackage[utf8]{inputenc}
\usepackage[T1]{fontenc}
\usepackage{graphicx}
\usepackage{caption}
\usepackage{subcaption}
\usepackage{setspace}
\usepackage{lineno}

\usepackage{soul}
\usepackage{todonotes}
\usepackage{tcolorbox}

\title{Electric Vehicles Limit Equitable Access to Essential Services During Blackouts}

\author[1,*]{Yamil Essus}
\author[1,2]{Benjamin Rachunok}

\affil[*]{yaessus@ncsu.edu}
\affil[1]{Department of Industrial and Systems Engineering, North Carolina State University, Raleigh, NC}
\affil[2]{Operations Research Program, North Carolina State University, Raleigh, NC}

\begin{abstract}
Electric vehicles (EVs) link mobility and electric power availability, posing a risk of making transportation unavailable during blackouts. We develop a computational framework to quantify the impact of EVs on mobility and access to services and find that existing access issues are exacerbated by EVs. Our results demonstrate that larger batteries reduce mobility constraints but their effectiveness is dependent on the geographic distribution of services and households. We explore the trade-offs between mobility and quality-of-life improvements presented by Vehicle-to-Grid technologies and the feasibility and trade-offs of public charging infrastructure as a solution to access inequalities. Equitable access to essential services (e.g. supermarkets, schools, parks, etc.) is the most important aspect of community resilience and our results show vehicle electrification can hinder access to essential services unless properly incorporated into policy and city-scale decision-making.

\end{abstract}
\begin{document}

\flushbottom
\maketitle
\thispagestyle{empty}

\spacing{1.5}


Vehicle electrification is a key component of Sustainable Development Goals yet the mass adoption of electric vehicles can lead to unintended consequences on community mobility during natural hazards. A growing body of research has identified positive and negative unintended consequences of the large-scale transition from gas to battery electric vehicles (BEVs). Beyond decarbonization, BEVs have the potential to positively impact air quality and environmental justice \cite{C1, C2, C3, C4, C5, C6} and to provide quality-of-life improvements through Vehicle-to-grid (V2G) technologies as vehicles can be used to power homes during outages \cite{F1, F2}.  However, there are growing questions about BEV carbon efficiency \cite{A1,A2} and mass-production feasibility \cite{B1, B2, B3, B4, B5} and BEVs also introduce challenges for integration into the grid\cite{D1}, create infrastructure vulnerabilities \cite{D2} and ties between efficiency and consumer behavior \cite{E3}.

BEVs also pose a challenge to owners during natural disasters \cite{G1, Almutairi_2022}. Electric vehicles couple transportation and electric power systems and this link poses a risk of power outages making transportation unavailable during disasters and limiting mobility for BEV owners during power outages. Because electric vehicles require periodic charging, the mobility potential of individual households is limited by their access to electricity and prolonged blackouts will constrain the ability of car-dependent households to access essential services. 

Due to the direct relation between driving distance to services and vehicle battery consumption, mobility will be impacted by geographic and technological factors. Geography determines the driving distance to essential services which will be translated to electricity consumption devoted to transportation. Technology determines the size and efficiency of electric vehicle batteries \cite{H2,H3} while V2G technologies introduce tradeoffs between quality of life improvements from powering household amenities and using electricity for mobility. 

The linkage between mobility, power availability and quality-of-life has broad implications for community resilience. Beyond the traditional focus on infrastructure functionality, equitable access to essential services is the most important aspect of community resilience \cite{H4}. In 2021, the transportation sector accounted for roughly 22\% of CO2 emissions \cite{iea2023,epa2023} and 55\% of transportation emissions in the US are attributed to gasoline vehicles \cite{usdoe}. Accordingly, battery electric vehicles are at the forefront of Sustainable Development Goals \cite{Winkler_Pearce_Nelson_Babacan_2023, E4, iea2021, iea2022, WBCSD} and the mass adoption of BEVs must be carefully managed to avoid negative impacts on community resilience.

We develop a computational modeling framework to quantify the impact of vehicle electrification on limited mobility and access to essential services in urban areas during prolonged blackouts. We define measures of access risk and evaluate how risk changes across large urban centers in the U.S. We find inequalities in access to essential services will be exacerbated by vehicle electrification during blackouts. We also find that urban areas with high population density are associated with lower levels of access risk whereas high car ownership rates correlate with higher access risk. Moreover, we test different electric vehicle technologies and find that increased battery size lowers access risk by increasing potential driving distance, however the impact of battery size is highly dependent on the geography of each city. Finally, we test how V2G creates a trade-off between access to services and use of household amenities and find that V2G disproportionately benefits access-rich households.

\subsection*{The link between mobility and electric power availability}

We model the impact of electrification on access to essential services and how access is mediated by technology and geography. We measure access risk at a household level for 16 major cities in the US. For each census block group in the city, we calculate the average distance to the nearest supermarket, school, and park and assume each household makes a given number of trips to each service per week (Methods). Driving distance is converted to weekly electricity consumption using standard BEV battery consumption rates \cite{ev_database} and we assume a fully charged battery at the start of the blackout to model available energy as a function of time. We define a measure of access risk which considers households at risk if two weeks of mobility activities consumes more than 50\% of their vehicle's energy reserves and assume a battery capacity of 60 kWh which is roughly equivalent to a Tesla Model Y (2022) and around the average capacity of available vehicles \cite{ev_database}. Starting with a full battery, the charge remaining after a two-week blackout represents the slack in mobility a household has after meeting basic mobility needs. We chose a two-week blackout because it lies in the range of recent events \cite{Macmillan_Wilson_Baik_Carvallo_Dubey_Holland_2023, hurricane_ida} and it allows us to explore the interactions between access risk and electric vehicles.

\begin{figure}[!th]
  \centering
  \includegraphics[width=\linewidth]{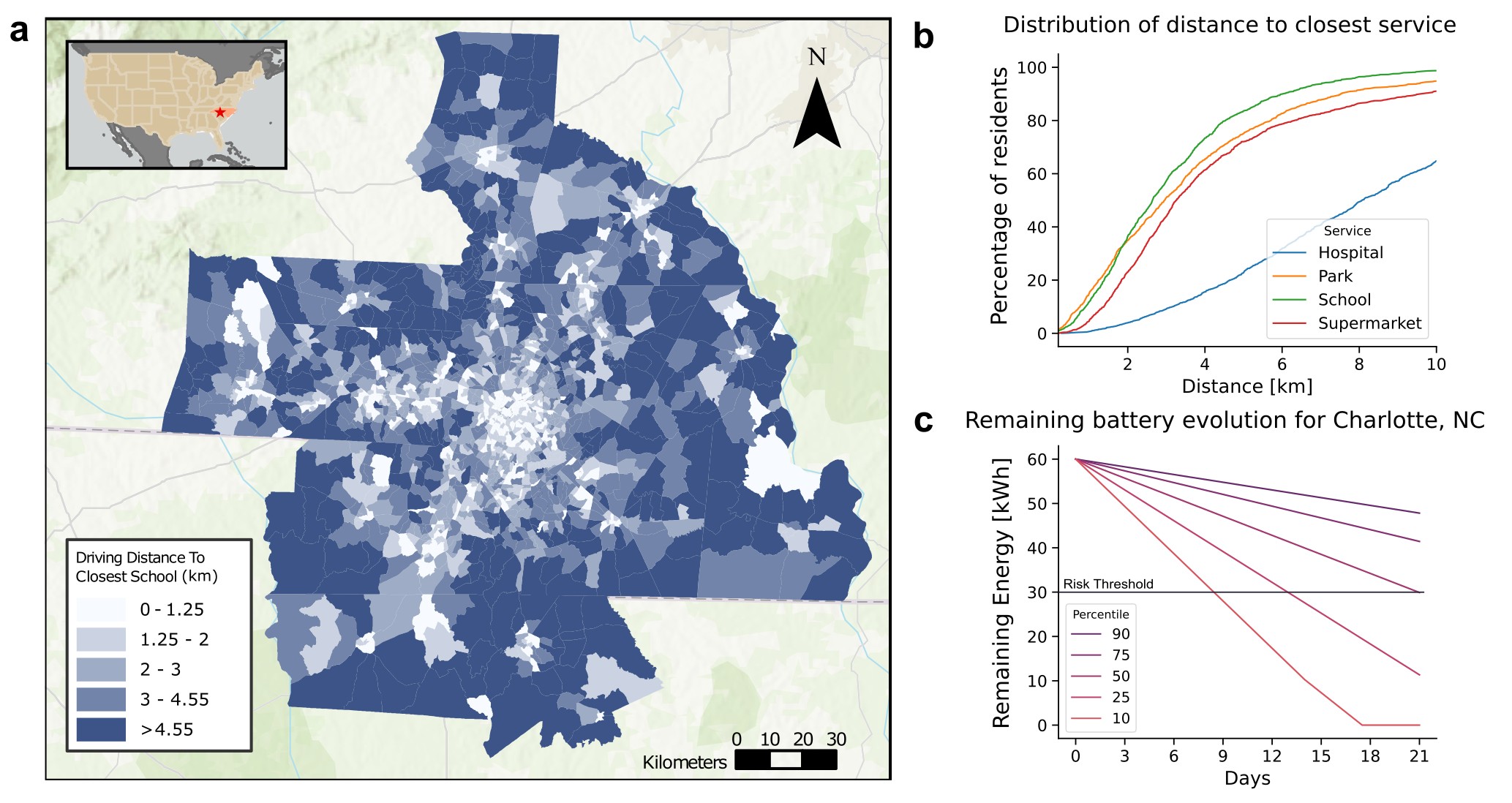}
  \caption{\textbf{The effect of geographic inequalities on access to essential services.} \textbf{(a)} Driving distance to the closest school in Charlotte, NC. \textbf{(b)} distribution of distance to closest services for the combined statistical area of Charlotte, NC.  and \textbf{(c)} Evolution of remaining battery available for 10th-90th percentile levels.  Map generated using TIGER files from the US Census Bureau, downloaded in April 2023. }
  \label{fig:fig1}
\end{figure}

We capture geographic inequality by computing driving distance to essential services including schools, supermarkets and parks. Figure \ref{fig:fig1}.a shows driving distance to the closest school for each census block group in Charlotte, NC and Figure \ref{fig:fig1}.b shows the distribution of driving distance for all services. Figure \ref{fig:fig1}.c shows the distribution of the State of Charge (SOC) over time. By evaluating the distribution of SOC we are able to identify census block groups at higher risk of losing access to essential services (households which are access-poor) as well as those with relatively little impacts (households which are access-rich). For example, it takes nine days for the bottom 10\% of the population to become at risk of losing access, whereas the top 10\% could withstand over 60 days without power before becoming at risk. By integrating characteristics of electric vehicles and driving distances at a census block group level we are able to assess the equity across block groups in terms of remaining battery resulting from existing inequalities in driving distance to services.


\begin{table}
    \centering
    \begin{tabular}{lcrr}
\toprule
              Name & Mean [75th - 25th] Remaining Battery &  Average Car Ownership &  Average Density \\
\midrule
           New York, NY &                51.26 [56.10 - 49.78] &                   0.52 &            12769 \\
     Salt Lake City, UT &                50.20 [53.99 - 49.74] &                   0.79 &             1987 \\
        Los Angeles, CA &                50.07 [54.34 - 49.25] &                   0.74 &             3940 \\
            Chicago, IL &                49.71 [54.91 - 47.85] &                   0.70 &             3354 \\
              Miami, FL &                47.85 [53.06 - 45.17] &                   0.73 &             3007 \\
           Portland, OR &                47.33 [54.10 - 47.59] &                   0.77 &             1945 \\
  Dallas-Fort Worth, TX &                45.72 [52.46 - 44.54] &                   0.77 &             1680 \\
        Kansas City, MO &                44.54 [52.44 - 41.89] &                   0.80 &              983 \\
            Houston, TX &                44.30 [51.50 - 42.47] &                   0.75 &             1835 \\
            Orlando, FL &                41.41 [48.97 - 37.12] &                   0.75 &             1317 \\
         Cincinnati, OH &                40.78 [50.17 - 34.27] &                   0.79 &             1001 \\
          St. Louis, LA &                39.14 [49.40 - 32.80] &                   0.80 &              868 \\
            Atlanta, GA &                38.74 [46.75 - 32.99] &                   0.77 &             1039 \\
          Nashville, TN &                36.41 [47.75 - 26.19] &                   0.81 &              763 \\
            Raleigh, NC &                36.11 [48.72 - 30.62] &                   0.79 &              788 \\
          Charlotte, NC &                35.70 [46.65 - 28.62] &                   0.79 &              728 \\
\bottomrule

\end{tabular}
    \caption{\textbf{Simulation Results}. For each Combined Statistical Area we include average remaining battery [kWh] as well as population weighted 25th and 75th percentile. Additionally, average population density [people/km2] and car ownership rates [cars/person on household] were obtained from open census data.}
    \label{tab:results}
\end{table}

\subsection*{Inequalities in access to essential services will be exacerbated by electric vehicles during blackouts}

We find that existing inequalities in geographical distribution of access to services will be exacerbated by the electrification of private cars because households further away from amenities will use more of their available energy on each trip. Accordingly, households in access-poor areas will be at risk of losing mobility due to this link between transportation and electricity. Average, 25th and 75th percentile values for the remaining battery of each city can be found in Table \ref{tab:results}. In the 16 cities tested, 
we find that more equitable access to essential services is associated with lower access risk. We used the Equally Distributed Equivalent (EDE) Index to quantify access inequalities. The EDE has been used to study inequality in urban settings with a focus on equitable distribution of burdens and resources \cite{H1} and can be interpreted as the value that would make an individual indifferent between the current distribution and complete equality at the level of the EDE. The EDE considers both the average value and the dispersion of a distribution, so that the EDE of two cities can be directly compared. Moreover, if two distributions share the same average, the EDE will penalize the one with higher dispersion.For example, the EDE of remaining battery of Portland is 43.6 kWh, which would increase if either the average remaining battery in Portland increases or if the inequality is reduced. The maximum EDE of remaining battery in Portland is limited to 60kWh and is achieved when every household has exactly 60kWh remaining. Details about the EDE metric used are included in the Methods section.

\begin{figure}[!t]
\centering
\includegraphics[width=\linewidth]{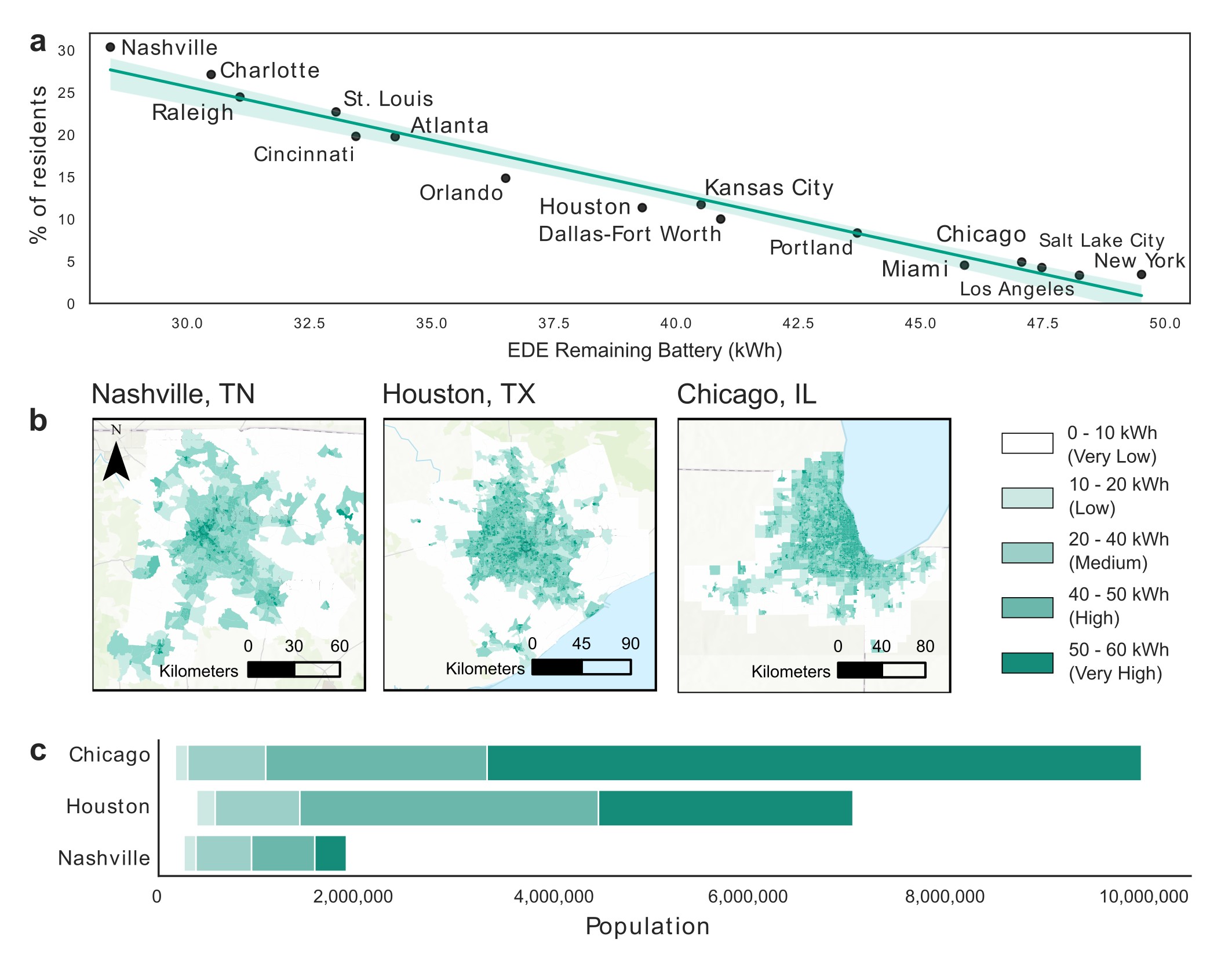}
\caption{\textbf{Geographic inequalities linked to access risk} \textbf{(a)} Equally Distributed Equivalent against percentage of residents experiencing access risk for Combined Statistical Areas in the U.S. \textbf{(b)} Available energy (kWh) after a two week blackout as a function of distance to amenities. \textbf{(c)} Population of the three cities shown in the map colored according to remaining battery levels. Maps were generated using TIGER files from the US Census Bureau database Redistricting 2022, downloaded in April 2023 
}
\label{fig:figure_2}
\end{figure}

We find a strong correlation between inequality in battery levels and access risk in each city we tested (Figure \ref{fig:figure_2}.a). For example,  Chicago has an EDE of 47.1 kWh and only 5\% of the residents would be at risk of becoming disconnected from essential services while Nashville has a EDE of 28.4 kWh and 30\% of residents experience access risk. 
Differences in geographic access between access-poor and access-rich households inside a city result in battery inequality that is captured by the EDE. The trend of greater EDE being associated with lower access risk demonstrates the link between the geographic aspects that drive consumption rates and our measure of risk as well as how BEVs exacerbate existing geographic inequalities during prolonged blackouts.

Access issues also demonstrate spatial trends. This is evidenced visually in the map in Figure \ref{fig:figure_2}.b and the population distribution in Figure \ref{fig:figure_2}.c colored according to the remaining battery of each census block after the blackout. We categorize each census block group according to estimated remaining energy using the intervals 0-10 kWh (Very Low), 10-20 kWh (Low), 20-40 kWh (Medium), 40-50 kWh (High), 50-60 kWh (Very High). Low remaining battery is associated with high risk, and high remaining battery with low risk. The same color palette was used for the maps and the population bar plots. In all three cities we identify regions of good access with short driving distances to services and sparse areas with poor connectivity that result in low battery for households located there. For example, Nashville has one center of very high remaining battery surrounded by large areas of medium to low remaining battery. On the other hand, Houston has a large number of small regions of high connectivity that end abruptly when moving away from the city center. Lastly, Chicago has a large area with very high remaining battery and only peripheral areas show low or very low levels. These results indicate that the geographic distribution of services and households greatly affects the access risk levels of an urban area.

We note that despite the fact geographic inequalities drive access risk, there is a significant number of people experiencing access risk in cities with relatively low inequality (Figure \ref{fig:figure_2}b). For example, both inequality and access risk are high in Nashville, TN, whereas Chicago, IL shows low inequality and low access risk, with over 80\% of the total population in the High or Very High level of battery remaining. Nonetheless, there is still over 500,000 people in Chicago experiencing access risk. These results indicate that the absolute and relative size of the population experiencing access risk are important for assessing how EVs effect mobility.

\subsection*{High population density and low car ownership rates are correlated with lower access risk}
\begin{figure}[!t]
\centering
\includegraphics[width=\linewidth]{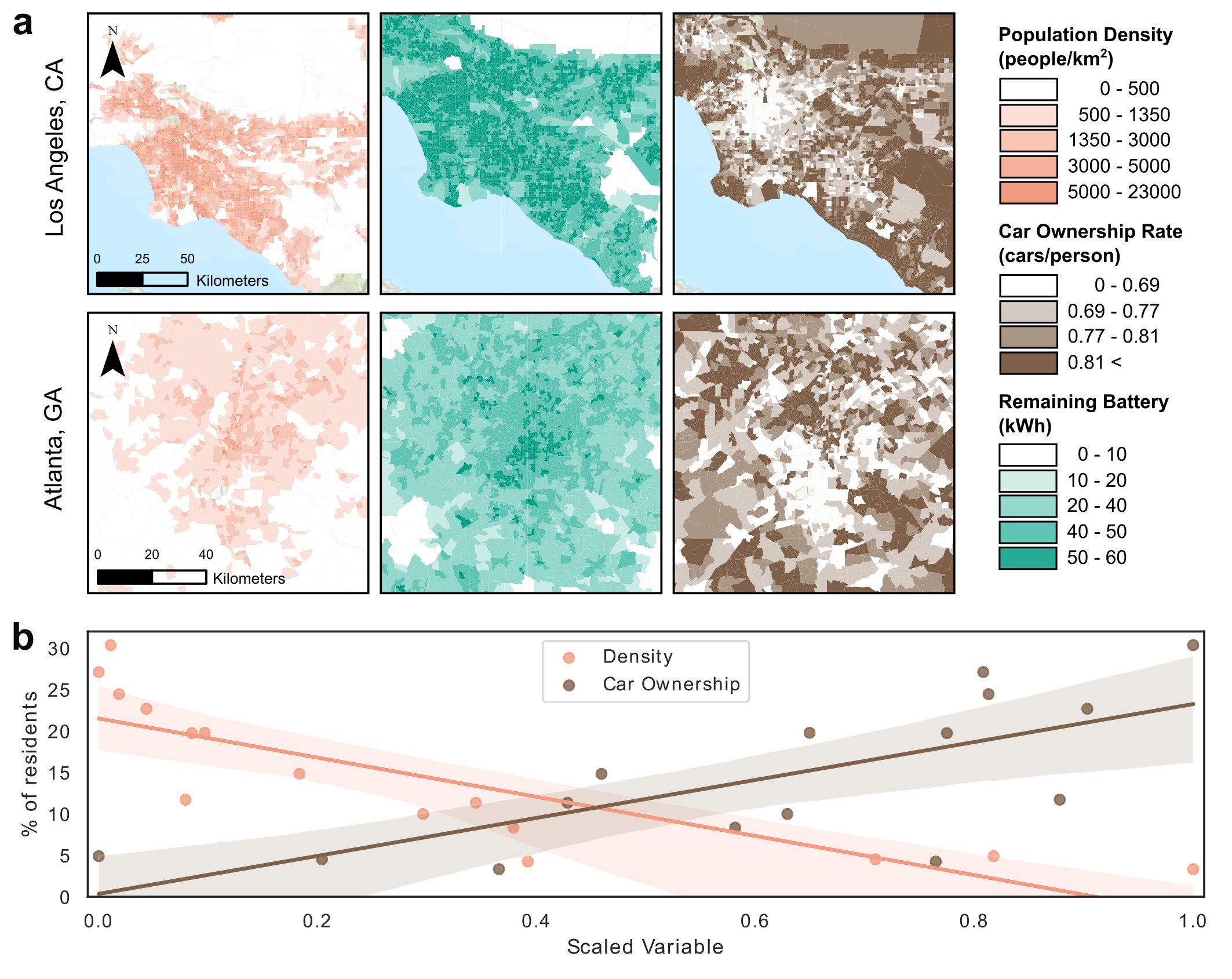}
\caption{\textbf{Urban form and access risk} \textbf{(a)} Log-minmax-scaled population density, battery SOC after a two week blackout and minmax-scaled car ownership rate for each census block group.  \textbf{(b)} Scaled population density and scaled car ownership plotted against percentage of population experiencing access risk for each of the CSA considered in the study. Values were aggregated from block group level using population-weighted average. New York City was excluded from this plot because it was considered an outlier. Maps were generated using TIGER files from the US Census Bureau database Redistricting 2022, downloaded in April 2023.}
\label{fig:figure_3}
\end{figure}

We find geographic inequality and access risk are associated with urban growth patterns. We quantify urban growth using population density and car ownership rates (Figures \ref{fig:figure_3}.a and \ref{fig:figure_3}.b) following research that has linked these metrics to social inequality in the context of sustainable development. For instance, inequalities in urban population density within a city can appear as a result of socio-political drivers, produce economic inequalities and propagate patterns of segregation \cite{Scheba_Turok_Visagie_2021}. Additionally, the dependence on cars of American households \cite{Moody_Farr_Papagelis_Keith_2021, Grengs_2010} is an obstacle for the development of sustainable cities because of its ties to income inequality \cite{Frederick_Gilderbloom_2018}. 

Population density is negatively correlated with access risk and car ownership rates are positively correlated with access risk across the U.S. Population density is inversely proportional to travel distance in the U.S. and is largely  reflective of other dimensions of disadvantage such as economic inequality \cite{Giuliano_Dargay_2006}. Additionally, car ownership rates quantify the perceived need for a personal car and the lack of reasonable public transport alternatives \cite{Frederick_Gilderbloom_2018}. Figure \ref{fig:figure_3}.b shows the aggregated value of density and car ownership variables for each city and the trend between them and access risk. Across cities we tested, we find that low population density and high car ownership correlate with high risk. These trends show the interaction between urban growth metrics and access risk and are evidence of how urban planning that incentivizes high population density and focuses on transportation policies that disincentivize car ownership can lead to lower access risk.

We also find that the trends between density, car ownership, and access risk are present within cities. Figure \ref{fig:figure_3}.a shows population density, average remaining battery and average car ownership rates for each census tract in Los Angeles, CA and Atlanta, GA. For example, the top and bottom 25\% most densely populated tracts in Atlanta have an average remaining battery of 45kWh [75\% battery level] and 23kWh [38\% battery level] respectively, while top and bottom 25\% by car ownership rates have 33kWh [55\% battery level] and 42kWh [70\% battery level] respectively, all against a city average of 37.5kWh [63\% battery level]. For Los Angeles, the top and bottom 25\% most densely populated tracts have an average remaining battery of 52kWh [87\%] and 40.5kWh [68\%] respectively, while top and bottom 25\% by car ownership rates have 45kWh [75\%] and 51kWh [85\%] respectively, all against a city average of 48.5kWh [82\%]. Associations between population density and decreased access risk as well as car ownership and increased access risk highlight the importance of identifying vulnerable populations even when the average risk for the entire area is relatively low.

\subsection*{The effectiveness of technological improvements on access risk depends on geography}

We find increased battery capacity or vehicle efficiency reduces access risk, but some areas are more sensitive to technological changes than others. We model improved vehicle energy efficiency or larger batteries as having the same effect. Greater efficiency reduces consumption per BEV trip and bigger batteries allow households to make more trips. We find that for all cities, technological upgrades reduce access risk but the magnitude of the decrease depends on the geographical distribution of households and services. We test the effects of reducing or increasing battery capacity on the portion of households experiencing access risk in each city. Three scenarios of initial battery capacity were tested: Low (40 kWh), Medium (60 kWh, base scenario) and High (80 kWh).

\begin{figure}[!t]
\centering
\includegraphics[width=\linewidth]{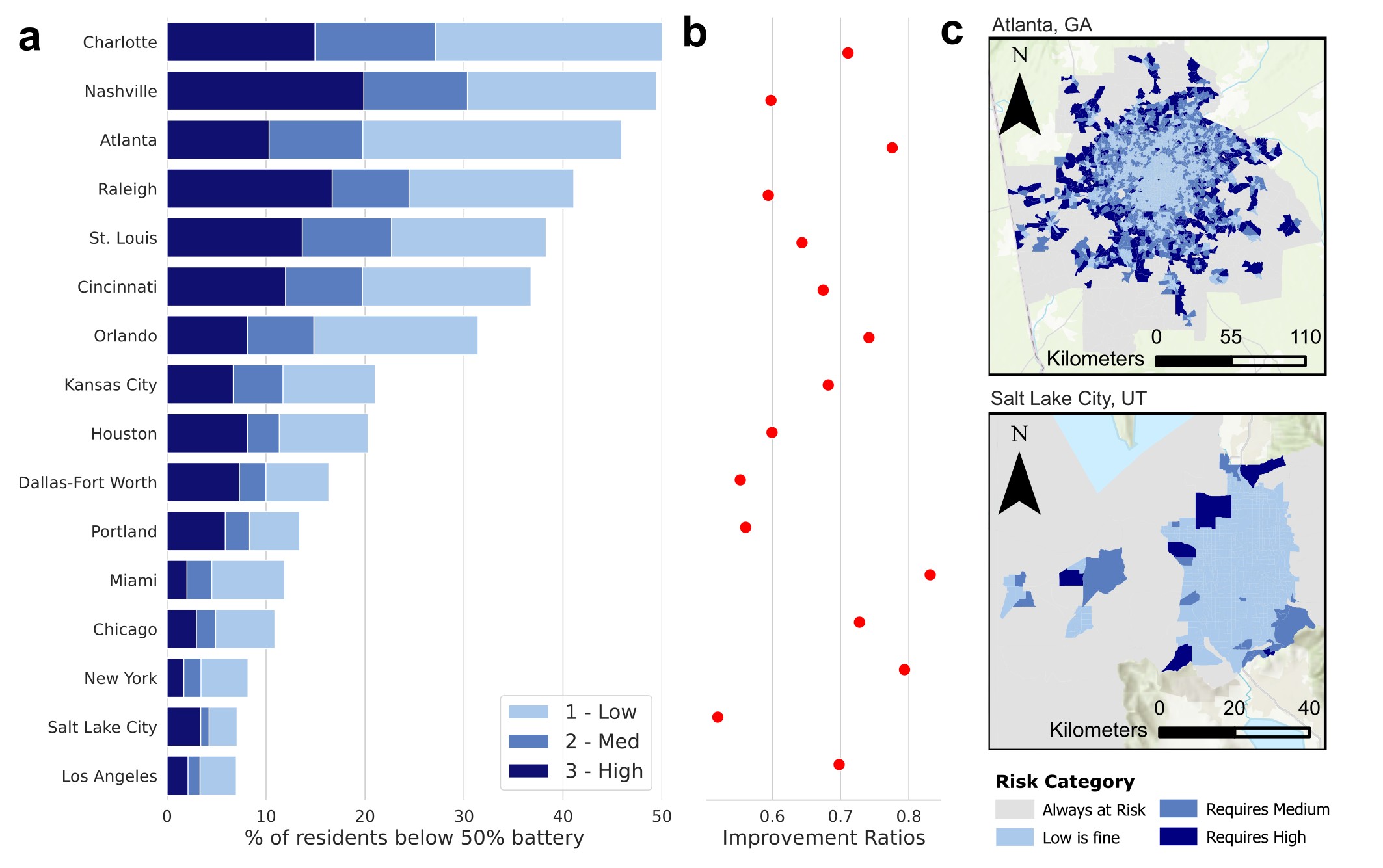}

\caption{\textbf{Impacts of battery size on access risk} \textbf{(a)} Percentage of residents experiencing access risk in a low, medium and high technology scenario (40, 60 and 80 kWh of initial battery charge respectively) for each CSA considered in this study. \textbf{(b)} Percentage of the portion of residents at risk in the low technology scenario that are not at risk in the high technology scenario for each CSA and \textbf{(c)} Geographic location of census block groups at risk that require  None/Medium/High battery capacity to evade the risk. Maps were generated using TIGER files from the US Census Bureau database Redistricting 2022, downloaded in April 2023}
\label{fig:figure_4}
\end{figure}

Figure \ref{fig:figure_4}.a shows the percentage of residents experiencing access risk for each scenario in each city. All cities follow the trend of risk being inversely proportional to battery size but geographic differences impact the change in access risk. For example, Orlando presents similar levels of access risk as Kansas City in the high battery scenario, but with low capacity batteries  10\% more of Orlando's households experience access risk compared to Kansas City's low battery scenario. This happens because both cities share a similar portion of very access-poor households that even in the high battery scenario are at risk, but there are more households that take advantage of the incremental increase in battery in Orlando. This means Kansas City's population is more concentrated into the extremes of access-rich households that are never at risk and access-poor households that require a bigger jump in battery to not experience risk. 

Figure \ref{fig:figure_4}.b highlights the portion of the households at risk in the low battery scenario that are not at risk in the high battery scenario for all cities tested. This value represents the maximum improvement in access risk that a change from 40 kWh to 80 kWh can achieve. For example, Atlanta’s geography allows for a decrease of 77\% in the percentage of people at risk when the technological scenario is favorable. On the other hand, Salt Lake City presents an improvement of around 50\% when comparing a Low battery vehicle scenario to a High battery vehicle scenario. This happens largely because of the geographical distribution of the population and how technological changes aligns with it. We can see in the map in Figure \ref{fig:figure_4}.c that Atlanta’s population that benefits from an increased battery capacity gradually increases as we move away from the urban center. In contrast, Salt Lake City’s map shows that most people living in the urban center are already in good standing when it comes to access risk, while people in the outskirts required an even larger increase in capacity to make it above the risk threshold.


\subsection*{The trade-offs between Vehicle-To-Grid and mobility}
Vehicle-to-grid (V2G) is a technology that enables energy to be returned to the grid or a house from a car battery \cite{v2g_report} and can provide significant quality of life improvements, but introduces a trade-off between mobility and household amenities. This trade-off is mediated by geography because it is determined by the rate at which households require electricity to access essential services. Using average household consumption rates, we explore the relation between mobility and household amenities for each geographic area.

\begin{figure}[h]
\centering
\includegraphics[width=.7\linewidth]{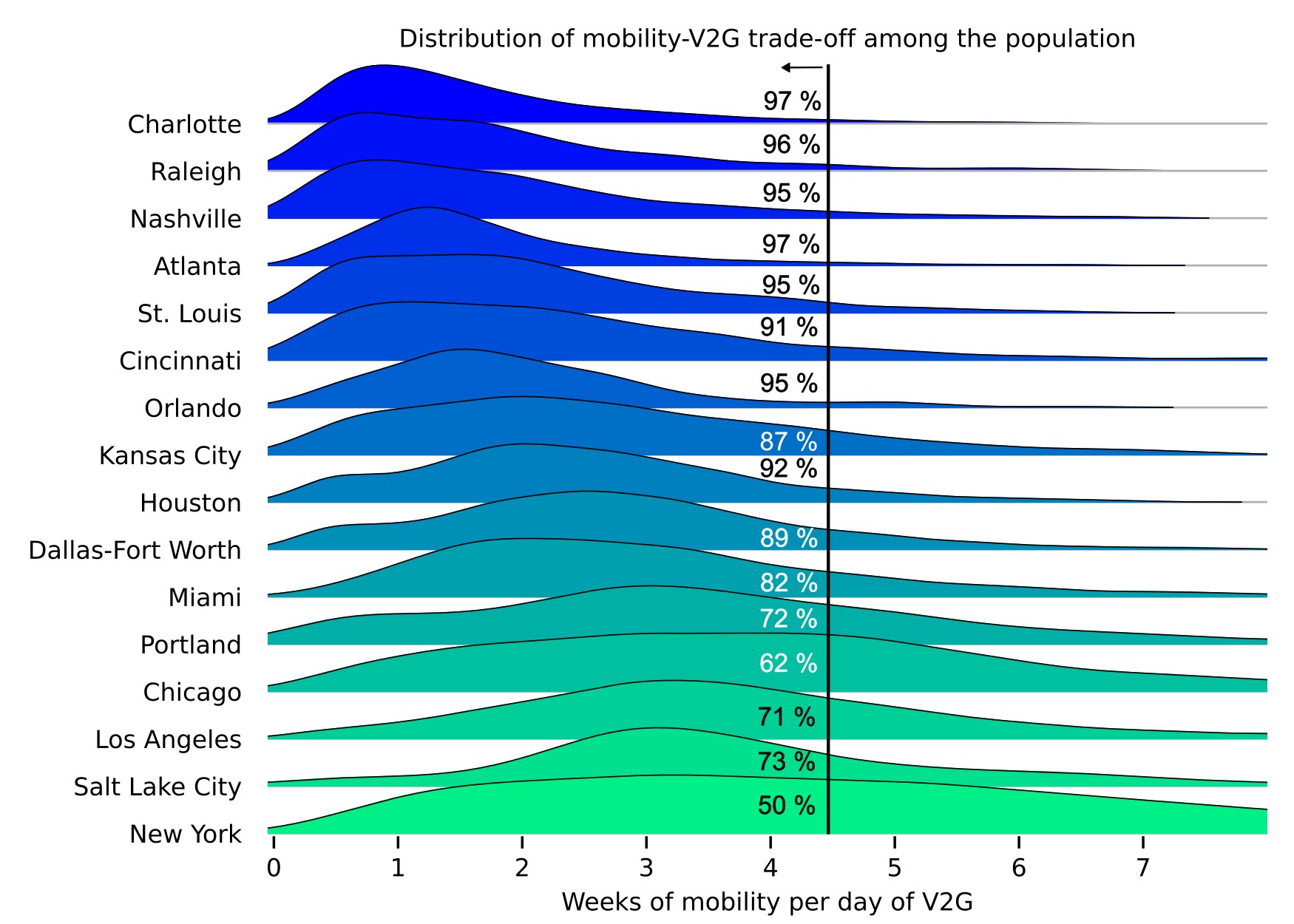}
\caption{\textbf{Vehicle-to-grid and Mobility Trade-offs.} Estimated distribution of the equivalent weeks of access to services per each day of average V2G consumption across the population of each city. Percentage labels show the portion of the population that has a mobility-V2G trade-off of at most 4.5 weeks of mobility per day of V2G, corresponding to the median of New York City. The figure was trimmed at 8 weeks of mobility per day of V2G. Each distribution is shown with its own normalized scale in the y-axis while all share the x-axis scale.
}
\label{fig:figure_5}
\end{figure}

Households with better access to services will sacrifice more days of mobility for each day of house amenities. This happens because better access leads to less energy consumption per trip therefore the same energy consumption on amenities for an access-rich household represents a larger number of days of mobility than for an access-poor household. Figure \ref{fig:figure_5} shows the distribution of this trade-off across the population of each city. We find that inequalities in access appear as differences in mobility-V2G trade-offs within each city. For instance, in Miami, FL some households have a trade-off of up to 7 weeks of mobility per day of V2G while the median household has a trade-off of 3.8 weeks per day of V2G. We also find differences across cities (Figure \ref{fig:figure_5}). The mobility-V2G trade-off experienced by the median household in NYC corresponds to around 4.5 weeks of mobility per day of V2G. We see that this level of V2G availability is better than the one experienced by 62\% of the population in Portland, OR and 97\% in Charlotte, NC. These results indicate that better access to services not only results in reduced access risk but also improves the usefulness of V2G.

Due to the high consumption of electricity for household amenities in comparison to vehicle batteries and mobility consumption, V2G technologies can only be realistically used by access-rich households. Estimated daily consumption in house amenities amounts to 13.3kWh (see Methods). Given the characteristics of the BEV considered in this study one day of V2G consumption amounts to weeks of mobility to access essential services. This means that only the households that are the least likely to experience access risk are will be likely to use this technology.

\section*{Discussion}


Using a computational framework informed by publicly available data, we show that access risk is present in urban environments with mass adoption of electric vehicles. Geographic inequality is translated into unequal electricity consumption rates which leads to an increased risk of losing access to essential services. Poor access gets exacerbated resulting in rapid consumption and low battery levels, thus, the inequality in remaining energy represents the weight of the lower end of the distribution of this risk for a given population. The geographic distribution of households and services is one of the main drivers of access risk. All else remaining equal, changes in geography will greatly impact access risk on almost all settings we tested. Our results indicate that policies that impact transportation systems, population density (such as district zoning laws) or car ownership rates (such as parking minimums and tax credit incentives) will all impact equitable disaster outcomes. This highlights how BEV mass adoption creates additional interdependencies in communities that must be included in policy decision-making.

A key result is that increases in battery capacity will reduce access risk but the effectiveness of larger batteries depends on geography. Electric vehicle technology is evolving quickly and there is rapid improvement in the battery capacity and efficiency of personal vehicles to overcome range limitations. These improvements can alleviate the burdens of access risk, but our results show that geographic disparities produce unequal benefits from increased battery sizes or efficiency gains. Because of this, technology-based public policies aiming to reduce access risk (e.g. economic incentives for more efficient or higher capacity vehicles) will have different outcomes depending on the geographic distribution of the population and services, and need to be evaluated for each region individually.  For example, the clean vehicle tax credit \cite{inflation_reduction_act_2022} subsidies EV purchases with up to \$ 7,500 in tax credits. Since the cost of the battery remains the most important cost of the manufacturing of BEVs \cite{DS2, DS3}, we estimate this translates into 7.5 kWh of additional battery assuming all the incentive is used towards increasing battery capacity (see Methods Section). This increase in battery would reduce households experiencing access risk by 9.2\% of total population in Charlotte, but only by 1.4\% in New York or 1.6\% in Chicago.



Another potential solution to EV-induced access risk is public charging. However, unless a separate electric distribution infrastructure is used, it will present the same limitations as residential charging during blackouts, with the added strain of long charging times. Additionally, as we find places with high density have the lowest access risk, citing charging stations presents a trade off between serving a large number of customers (placement in high-density locations) and those with high access risk (generally low-density locations). This same trade-off will be present to electricity utilities allocating resources after power outages. The objectives of returning power for the greatest number of people, and doing so in a way to not exacerbate access risk will trade off against eachother.

In addition to infrastructure and transportation, access risk impacts other dimensions of community resilience \cite{Cutter_Barnes_Berry_Burton_Evans_Tate_Webb_2008}. As a vulnerability of the transportation sector, access risk prevents the normal operation of businesses during blackouts, impacting economic resilience\cite{Rose_2004}. Additionally, social resilience can be affected by access risk due to the ties between risk and existing geographic inequality. Because inequality exists across and within cities, associations between historically marginalized communities and geographic inequality inside an urban area will exacerbate social disparities.

One way to adapt to access risk in the event of a long blackout is to directly supply resources to vulnerable communities. During disruptive events, local governments often put plans in place to help distribute essential goods, such as food or water, that may be hard to obtain. Existing ways to distribute aid can be altered to also support reducing access risk problems. This comes with potential negative consequences which is that this policies increase the pressure on support systems because current access issues are exacerbated by vehicle electrification. Additionally, households facing access risk will be more dependent on external support networks that will meet increased demand which needs to be accounted for.


Our work aims to identify large scale interactions in a radically different paradigm of individual mobility --assuming full electrification of personal vehicles-- which requires assumptions about human behavior and available technology. For instance, the number of visits to each service will vary across households, but a fixed number of trips still exposes the general relation between geography and risk. Similarly, we considered 50\% of the energy remaining as the cutoff for access risk, and changing that threshold up and down will increase or decrease the populations at risk.  However, range anxiety has been repeatedly identified in the context of electric vehicle adoption \cite{DS1} and is likely to have an effect on how this risk is perceived. Additionally, daily commutes, evacuation attempts and leisure activities will represent important loads for the energy reserves that have not been accounted for in our study. Lastly, in order to estimate the V2G thresholds experienced by each household, we assumed a fixed quantity of household electricity consumption (Methods). We also note that access risk will still exist with only Internal Combustion Engine Vehicles (ICEVs). Ongoing work is aimed at understanding the specific change in access risk between ICEVs and BEVs. However, an important takeway from our work is that access risk will still be present under mass adoption of electric vehicles --regardless of the access risk present with ICEV ownership.



\section*{Methods}
\label{sec:methods}

\subsection*{Model Overview}
We compute access risk at a census block group level based on weekly electricity consumption required to access essential services which is a function of geography. Weekly electricity consumption on mobility is estimated from driving distance to the closest set of services and assuming average vehicle efficiency (kWh/km) and uniform travel patters. We assumed complete adoption of BEV for residential mobility and travel patterns for all households consisting in two round trips to the closest supermarket, five to the closest school and three to the closest park on a weekly basis. The number of trips per week to each service was chosen to reflect basic mobility patterns to access essential services. The number of trips has a linear relationship with electricity consumption and a fixed number of trips allow us to connect geographic inequality with access risk. Additionally, no commuting was considered. Total distance driven for all services was computed for each census block group and was multiplied by the nominal efficiency of the car to obtain weekly battery consumption rates. No geographical or climatic variables were considered to affect vehicle efficiency.

\subsection*{Data}
Combined Statistical Areas (CSA) were used to identify the limits of each city. We chose a sample from the top 50 CSA by population according to the numbers reported by the U.S. Census bureau as of July 1, 2022. We selected most of the ten most populated areas and a sparse sample of the rest aiming for variety in population size and geographic location. Each combined statistical area is identified by a set of counties. For each county, locations of all schools, supermarkets, parks and hospitals were obtained using the Overpass API from OpenStreetMaps. Open Source Routing Machine \cite{project_osrm} was used to compute the closest driving distance to each type of amenity for each Census Block Group in each CSA. The centroid of each Census Block Group was used to compute this distance. Additionally, the 2021 5-Year survey data from the American Community Survey was used to obtain population estimates and car ownership rates per person per household for each census tract. Population density was estimated using total land area information included in the TIGER files. Aggregation of population density and car ownership rates per CSA was done weighting each census tract by its population in order to account for census tracts that represented a small percentage of the population.

\subsection*{Inequality metric}
The Equally Distributed Equivalent (EDE) metric was used to quantify and compare inequality in the distributions of access for each geographical area. Sheriff and Maguire \cite{M1} introduced the benefits of Equally distributed equivalents for quantifying risk and burden inequality in addition to resources. In particular, the EDE metrics are beneficial because they capture level and dispersion of the distribution, in contrast to popular metrics like the Gini index which is only concerned with dispersion. The Kolm–Pollak EDE for a distribution $X$ is defined according to Equation \ref{eq:ede} and has been used for measuring inequalities in urban systems \cite{H1}. 

\begin{equation}
    \Xi(X) = -\frac{1}{\kappa} ln\left [ \frac{1}{N} \sum_{i=1}^{N} e^{\kappa x_i} \right]
    \label{eq:ede}
\end{equation}

Where $\Xi$ is the Equally distributed equivalent, $N$ is the sample size and $x_i$ is the $i^th$ value in the distribution of $X$. The inequality aversion parameter $\kappa$ is input by the user and represents a weight for dispersion in the distribution. Atkinson \cite{M2} propose to calculate $\kappa$ according to Equation \ref{eq:kappa}, where $\epsilon$ is a user input parameter that whose absolute value is typically between 1 and 2 and its sign determines the nature of the distribution. Positive values are used to quantify dispersion in desirable quantities and negative values are used for burdens. The value of $\epsilon$ used in this work is 1.75.

\begin{equation}
    \kappa = \frac{\sum_{i=1}^{N}x_i}{\sum_{i=1}^{N}x_i^2} \epsilon
    \label{eq:kappa}
\end{equation}

\subsection*{Technology}
Base scenario electric vehicle battery capacity and efficiency were chosen to correspond to a Tesla Model Y (2022) according to the EV-Database website \cite{ev_database}. In particular, this means 57.5 kWh of usable battery (approximated to 60kWh to facilitate the generation of scenarios with increased/reduced battery) and a range of 350km resulting in an average efficiency of 5.83 km/kWh. The database was also used for estimating a cost per kWh of additional battery in the current market. We found that below \$80,000 in price, vehicles showed a trend of \$1,000 per additional kWh. Above that price, additional increases were not translated to increase capacity.

Household weekly electricity consumption was estimated using yearly residential consumption averages by end-use from the Energy Information Administration of the U.S\cite{eia_report}. The end-uses considered for daily V2G load were water heating, air conditioning, lighting and clothes dryer and the average consumption values were 28.14 kWh, 34.84 kWh, 21.24 kWh and 9.34 kWh per week respectively.

\bibliography{main}

\section*{Acknowledgements (not compulsory)}
Y.E. would like to thank and acknowledge funding from the Edward P. Fitts Doctoral Scholarship.

\section*{Author contributions statement}
Y.E. and B.R. conceived the idea; Y.E. conducted the analysis and experiments; Y.E. and B.R. drafted, reviewed and edited the manuscript. B.R. advised and funded.


\end{document}